\documentclass[a4paper,12pt]{article}
\usepackage{amssymb,cite}
\usepackage{amsmath}
\usepackage{amsthm}
\usepackage{latexsym}
\def\beq{\begin{eqnarray}}
\def\eeq{\end{eqnarray}}
\def\bsp{\begin{split}}
\def\esp{\end{split}}

\def\dx{{\bf dx}}
\def\dt{{\bf dt}}
\def\dy{{\bf dy}}
\def\dz{{\bf dz}}
\def\dw{{\bf dw}}

\newcommand{\mb}[1]{{\mathbb #1}}

\newcommand{\mbold}[1]{\mbox{\boldmath${#1}$}}
\newcommand{\be}{\begin{eqnarray}}
\newcommand{\ee}{\end{eqnarray}}
\begin{document}
\title{\Large\textbf{Essential Constants for Spatially Homogeneous Ricci-flat manifolds of
dimension 4+1}}
\author{\textbf{T. Christodoulakis}\thanks{tchris@cc.uoa.gr}\\
 University of Athens, Physics Department\\
Nuclear \& Particle Physics Section\\
Panepistimioupolis, Ilisia GR 157--71, Athens, Hellas
\vspace{.5cm} \\
\textbf{S. Hervik}\thanks{S.Hervik@damtp.cam.ac.uk} \\
 DAMTP, Centre for Mathematical Sciences,\\ Cambridge University, \\
Wilberforce Rd., Cambridge CB3 0WA, United Kingdom
\vspace{.5cm} \\
\textbf{G.O. Papadopoulos}\thanks{gpapado@cc.uoa.gr}\\
 University of Athens, Physics Department\\
Nuclear \& Particle Physics Section\\
Panepistimioupolis, Ilisia GR 157--71, Athens, Hellas}
\date{}
\maketitle
\numberwithin{equation}{section}
\begin{abstract}
The present work considers (4+1)-dimensional spatially homogeneous vacuum
cosmological models. Exact solutions --- some already
existing in the literature, and others believed to be new
--- are exhibited. Some of them are the most general for
the corresponding Lie group with which each homogeneous slice is
endowed, and some others are quite general. The characterization
``general'' is given based on the counting of the essential
constants, the line-element of each model must contain; indeed,
this is the basic contribution of the work. We give two
dif\mbox{}ferent ways of calculating the number of essential
constants for the simply transitive spatially homogeneous
(4+1)-dimensional models. The first uses the initial value
theorem; the second uses, through Peano's theorem, the so-called
time-dependent automorphism inducing dif\mbox{}feomorphisms.
\end{abstract}
\newpage
\section{Introduction}
Since the dawn of General Relativity people have been interested in
finding exact solutions to Einstein's Field equations\footnote{See
e.g. \cite{Kramer} for the known exact solutions in 3+1 dimensions.}. However, due to
the fairly complicated nature of the field equations, we usually impose
symmetries in order to make the field equations more tractable. Some of
the most successful schemes of symmetry reductions are the so-called
Bianchi models in (3+1)-dimensional cosmology\cite{bianchi,Ellis,EM}. Here, in this paper we
will consider their (4+1)-dimensional counterparts \cite{sig4+1,JPA}.

The study of higher-dimensional models has -- especially since the
advent of String Theory \cite{polchinski,green} -- become increasingly
popular in the recent years. For example,
exact solutions like plane-wave spacetimes have been in focus for the last
few years because they admit supersymmetry and they provide with an
exactly soluble string background. Plane-wave spacetimes are
some of the solutions  of the models considered
here. More specifically, we will consider (4+1)-dimensional spatially
homogeneous spacetimes which are solutions of the vacuum field
equations. Equivalently, we will consider Ricci-flat spacetimes;
i.e. spacetimes obeying
\beq
R_{ab}=0,
\eeq
which admit a group acting simply transitively on the spatial hypersurfaces.
The question we are addressing is ``How large is the set of
Ricci-flat spacetimes within the set of models considered?'', or
equivalently ``How
many parameters are necessary, in principle, to specify a solution of
this equation provided that they are spatially homogeneous?''. The
answer to this question turns out to be that 11 parameters are needed
to be specified to give a solution for the most general classes.

Many interesting phenomena are related to this issue. For example, a
by-product of
our analysis is that we are able to  determine which are the most
general vacuum models within the class of spatially homogeneous
models. In
(3+1)-dimensional cosmology, the most general simply connected Bianchi
vacuum models, namely type VIII, IX, and the exceptional model
VI$^*_{-1/9}$ \cite{Ellis}\footnote{Strictly speaking, it depends how
one counts. If we include the group parameter as an essential
constant, then types VI$_h$ and VII$_h$ are equally general.}, are all chaotic in the initial singular
regime \cite{Belinsky:1970ew,Spokoiny:1981fs,jb2,jb3,cb,dem,DH2000,dh2,hbc,hhw}. In 4+1
cosmology one might wonder if the same is the case\footnote{This was
  recently addressed for some of the models in \cite{dBPS}.}. We will also give
some exact solutions, thereby providing some examples of
spacetimes of each class. In some cases, the entire family is known
explicitly; in
others only a few, or even none, are known. However, some of these
special solutions have some interesting properties --- like
self-similarity --- which may be important in the late-time behaviour for
more general solutions (see e.g. \cite{Ellis}). As an explicit example
of this,
the plane-wave solutions -- which will be discussed later -- were shown
in \cite{HKL} to be the attractors within their class of models.

The paper is organized as follows. Next, we introduce the
automorphism group and see how it is related to coordinate
transformations, or the gauge freedom in general relativity (see
e.g. \cite{RE,JMP}). Then, in section \ref{counting}, we present
our main results, namely the counting of the essential constants
for the simply transitive, spatially homogeneous models of
dimension 4+1. In section \ref{exact} we present some exact
solutions before we conclude in section \ref{conclusion}.

\section{The r\^ole of the Automorphism group}
Let us first exhibit some basic assumptions that lie in the
foundation of our work.

Spacetime is assumed to be the pair $(\mathcal{M}, g)$ where
$\mathcal{M}$ is a 5-dimensional, Hausdorf\mbox{}f, connected,
time-oriented and $C^{\infty}$ manifold, and $g$ is a $(0,2)$ tensor
field, globally defined, $C^{\infty}$, non-degenerate and
Lorentzian (i.e. it has signature $(-,+,+,+,+)$). In the spirit of
4+1 analysis, we foliate the
entire spacetime like $\mathcal{M}=\mb{R}\times \Sigma_{t}$, where the
4-dimensional orientable submanifolds $\Sigma_{t}$ (\emph{surfaces
of simultaneity}) are space-like surfaces of constant time. The
assumption of spatial homogeneity corresponds to imposing the
action of a symmetry group of transformations $G$ upon the
manifolds $\Sigma_{t}$. Usually, the group $G$ is not only
continuous, but also a Lie group ---thus denoted by $G_r$, where
$r$ is the dimension of the space of its parameters. Avoiding the
details on these issues ---these matters can be easily found in
a standard reference, see e.g. \cite{Eisenhart,Petrov}--- we
simply state that spatially homogeneous models with a simply
transitive action of the group $G_{4}$ are described (apart from the
topology of $\Sigma_{t}$ which we will assume is simply connected) by an invariant basis of one-forms
${\mbold\omega}^{\alpha}=\sigma^{\alpha}_{i}(x)\dx^{i}$ (their Lie derivative with respect
to the generators of the Lie group $G_{4}$ ~$\xi^{i}_{\alpha}(x)$,
are zero). There is also the case of spatially homogeneous models
in which the group acts multiply transitively (the number of
generators is more than 4 and there does not exist a proper
invariant subgroup of dimension 4 acting transitively). The multiply transitive cases,
which in dimension 4+1 are 5 in total \cite{sig4+1}, will
not be considered here. More generally, a Lie Group,
$G_{r}$, is said to act transitively if the following requirements are
satisfied:
\begin{enumerate}
\item[(i)] $r\geq \textrm{dimension of~} \Sigma_{t}(=4)$
\item[(ii)] the rank of the matrix constructed by the generators
(seen as vector fields) are everywhere equal to the $\textrm{dimension
of~}\Sigma_{t}(=4)$.
\end{enumerate}
Geometrically, the above requirements imply that two dif\mbox{}ferent
points in a given domain of $\Sigma_{t}$ can be interchanged by a Lie
group transformation. Simply transitive action,
which is our concern  in this paper,
corresponds to the case $r=\textrm{dimension of~}\Sigma_{t}(=4)$.
In this case we note that the most
general line element, manifestly invariant under the action of the
group, takes the form (in an appropriate coordinate
system):\footnote{Greek indices label the invariant one-forms while
the Latin indices run over the spatial coordinates (i.e. from 1-4).}
\be
\label{lineelement}
ds^{2}=(N^{\alpha}(t)N_{\alpha}(t)-N^{2}(t))\dt^{2}+2N_{\alpha}(t)\sigma^{\alpha}_{i}(x)
\dt\dx^{i}+\gamma_{\alpha\beta}(t)\sigma^{\alpha}_{i}(x)\sigma^{\beta}_{j}(x)\dx^{i}\dx^{j},
\ee with
\be \label{rotation}
\sigma^{\alpha}_{i,j}(x)-\sigma^{\alpha}_{j,i}(x)=2C^{\alpha}_{\mu\nu}
\sigma^{\mu}_{i}(x)\sigma^{\nu}_{j}(x),
\ee
where
$\gamma_{\alpha\beta}(t)$ is the metric induced on the surfaces
$\Sigma_{t}$ (and thus constant on them); $N(t)$ is the lapse
function; $N_{\alpha}(t)$ is the shift vector
($N^{\alpha}(t)=\gamma^{\alpha\beta}(t)N_{\beta}(t)$,
$\gamma^{\alpha\beta}(t)$ being the inverse of
$\gamma_{\alpha\beta}(t)$); and $C^{\alpha}_{\mu\nu}$ are the
structure constants of the corresponding (closed) Lie algebra.
 In 4 dimensions, there are 30 closed, real, Lie algebras \cite{Patera1,Patera2}.

At this point, a question arises; is there any particular class of
General Coordinate Transformations (G.C.T.s) which can serve to
simplify the form of the line element and thus also  Einstein's
Field Equations (E.F.E.s)? The answer is positive and a thorough
investigation of this problem and its consequences is given in
\cite{JMP}; indeed, not only is there a class of G.C.T.s which
preserves the manifest spatial homogeneity of the line element
(\ref{lineelement}), but it also forms a continuous (and virtually,
Lie) group. This group is closely related to the symmetries of the
symmetry Lie group $G_{r}$; it is its automorphism group.

In the spirit of the 4+1 analysis we consider, apart from the time
reparameterization, the following G.C.T.'s :
\begin{align} \label{transformations}
t\rightarrow \widetilde{t}=t&\Leftrightarrow t=\widetilde{t}\nonumber \\
x^{i}\rightarrow \widetilde{x}^{i}=g^{i}(t,x^{j})&\Leftrightarrow
x^{i}=f^{i}(t,\widetilde{x}^{j}).
\end{align}
After insertion of (\ref{transformations}) into
(\ref{lineelement}), the wish to preserve the manifest homogeneity
of the latter leads, in a first step, to the allocations:
\begin{subequations}
\begin{align}
\frac{\partial f^{i}}{\partial t}&=\sigma^{i}_{\alpha}(f)P^{\alpha}(t,\widetilde{x})\\
\frac{\partial f^{i}}{\partial
\widetilde{x}^{j}}&=\sigma^{i}_{\alpha}(f)
\Lambda^{\alpha}_{\beta}(t,\widetilde{x})
\sigma^{\beta}_{j}(\widetilde{x}),
\end{align}
\end{subequations}
and, consequently,the definitions
\begin{subequations}
\begin{align}
\widetilde{N}(t)&=N(t)\\
\widetilde{N}^{\alpha}(t)&=S^{\alpha}_{\beta}(t)(N^{\beta}(t)+
P^{\beta}(t,\widetilde{x}))\\
\widetilde{\gamma}_{\alpha\beta}(t)&=\Lambda^{\mu}_{\alpha}(t,\widetilde{x})
\Lambda^{\nu}_{\beta}(t,\widetilde{x}) \gamma_{\mu\nu}(t),
\end{align}
\end{subequations}
where
$N^{\alpha}(t,\widetilde{x})=\gamma^{\alpha\beta}(t)N_{\beta}(t,\widetilde{x})$
with $\sigma^{i}_{\alpha}(x)$ being the inverses of
$\sigma^{\alpha}_{i}(x)$ ---quantities which exist in the simply
transitive cases. In order for the transformations
(\ref{transformations}) to have a well defined non-trivial action,
it is pertinent for the quantities $\Lambda^{\alpha}_{\beta}$ and
$P^{\alpha}$ to be space independent. So,
\begin{subequations} \label{allocations}
\begin{align}
\frac{\partial f^{i}}{\partial
t}&=\sigma^{i}_{\alpha}(f)P^{\alpha}(t)\\
\frac{\partial
f^{i}}{\partial\widetilde{x}^{j}}&=\sigma^{i}_{\alpha}(f)
\Lambda^{\alpha}_{\beta}(t) \sigma^{\beta}_{j}(\widetilde{x}),
\end{align}
\end{subequations}
and therefore
\begin{subequations} \label{newsetofvariables}
\begin{align}
\widetilde{N}(t)&=N(t)\\
\widetilde{N}^{\alpha}(t)&=S^{\alpha}_{\beta}(t)(N^{\beta}(t)+P^{\beta}(t))\\
\widetilde{\gamma}_{\alpha\beta}(t)&=\Lambda^{\mu}_{\alpha}(t)
\Lambda^{\nu}_{\beta}(t)\gamma_{\mu\nu}(t).
\end{align}
\end{subequations}

Thus (\ref{allocations}) instead of being allocations, turn into a
set of highly non-linear partial dif\mbox{}ferential equations.
Integrability conditions for this system, i.e. Frobenious' Theorem,
results in the system (the dot, whenever used, denotes
dif\mbox{}ferentiation with respect to time):
\begin{subequations} \label{Aut}
\begin{align}
C^{\beta}_{\mu\nu}\Lambda^{\alpha}_{\beta}(t)&=C^{\alpha}_{\kappa\lambda}
\Lambda^{\kappa}_{\mu}(t)\Lambda^{\lambda}_{\nu}(t)\\
\frac{1}{2}\dot{\Lambda}^{\alpha}_{\beta}(t)&=C^{\alpha}_{\mu\nu}P^{\mu}(t)
\Lambda^{\nu}_{\beta}(t),
\end{align}
\end{subequations}
and \emph{``Time-Dependent Automorphism Inducing
Dif\mbox{}feomorphisms''} (A.I.D.s) emerge. The automorphisms of a
Lie group $G_{r}$ form a continuous group. Those members of the
group which are continuously connected to the identity element,
form a Lie group as well --- even though the topology of the
latter might be dif\mbox{}ferent from that of the former. If one
considers parametric families of the automorphic matrices,
characterized by the parameters $\tau^{i}$,
$\Lambda^{\alpha}_{\beta}(t;\tau^{i})$, and defines:
\begin{subequations}
\begin{align}
\Lambda^{\alpha}_{\beta}(t;\tau^{i})\Big|_{\tau^{i}=0}&=\delta^{\alpha}_{\beta}\\
\frac{d\Lambda^{\alpha}_{\beta}(t;\tau^{i})}{d\tau^{i}}\Big|_{\tau^{j\neq
i}=0}&= \lambda^{\alpha}_{\beta(i)}
\end{align}
\end{subequations}
where $\lambda^{\alpha}_{\beta(i)}$ are the generators with
respect to the parameter $\tau^{i}$ of the Lie algebra of the
automorphism group, then from the first of (\ref{Aut}), after a
dif\mbox{}ferentiation with respect to $\tau^{i}$, one gets
\be\label{generators}
\lambda^{\alpha}_{\beta(i)}C^{\beta}_{\mu\nu}=\lambda^{\rho}_{\mu(i)}
C^{\alpha}_{\rho\nu}+\lambda^{\rho}_{\nu(i)}C^{\alpha}_{\mu\rho}.
\ee For an extensive treatment on these issues see \cite{CMP},
while for the relation and usage of these generators with
conditional symmetries, see \cite{Kuchar,CQGTYPEI}.

In the $n+1$ decomposition of the spacetime (here $n=4$), the
E.F.E.s in vacuum assume the form:
\begin{subequations} \label{EinsteinEq}
\begin{align}
E^{0}_{0}&=K^{\alpha}_{\beta}K^{\beta}_{\alpha}-K^{2}+R=0
\label{quadraticconstraint}\\
E^{0}_{\alpha}&=K^{\mu}_{\nu}C^{\nu}_{\alpha\mu}-K^{\mu}_{\alpha}
C^{\nu}_{\mu\nu}=0\label{linearconstraint}\\
E^{\alpha}_{\beta}&=\dot{K}^{\alpha}_{\beta}-NKK^{\alpha}_{\beta}+NR^{\alpha}_{\beta}+
2N^{\rho}(K^{\alpha}_{\nu}C^{\nu}_{\beta\rho}-K^{\nu}_{\beta}C^{\alpha}_{\nu\rho})=0
\label{equationofmotion}
\end{align}
\end{subequations}
with
\begin{subequations}
\begin{align}
K^{\alpha}_{\beta}(t)&=\gamma^{\alpha\rho}(t)K_{\rho\beta}(t)\\
R^{\alpha}_{\beta}(t)&=\gamma^{\alpha\rho}(t)R_{\rho\beta}(t)\\
K_{\alpha\beta}(t)&=-\frac{1}{2N(t)}(\dot{\gamma}_{\alpha\beta}(t)+
2\gamma_{\alpha\nu}(t)
C^{\nu}_{\beta\rho}N^{\rho}(t)+2\gamma_{\beta\nu}(t)C^{\nu}_{\alpha\rho}N^{\rho}(t))\\
\begin{split} \label{curvature}
R_{\alpha\beta}(t)&=C^{\kappa}_{\sigma\tau}
C^{\lambda}_{\mu\nu}\gamma_{\alpha\kappa}(t)
\gamma_{\beta\lambda}(t)\gamma^{\sigma\nu}(t)\gamma^{\tau\mu}(t)+
2C^{\lambda}_{\alpha\kappa}C^{\kappa}_{\beta\lambda}+
2C^{\mu}_{\alpha\kappa}C^{\nu}_{\beta\lambda}\gamma_{\mu\nu}(t)
\gamma^{\kappa\lambda}(t)\\
&+2C^{\lambda}_{\alpha\kappa}C^{\mu}_{\mu\nu}\gamma_{\beta\lambda}(t)
\gamma^{\kappa\nu}(t)+
2C^{\lambda}_{\beta\kappa}C^{\mu}_{\mu\nu}\gamma_{\alpha\lambda}(t)
\gamma^{\kappa\nu}(t).
\end{split}
\end{align}
\end{subequations}
 Since G.C.T.s are covariances of the E.F.E.s, the same form of
equations (\ref{EinsteinEq}) holds for the transformed quantities:
\begin{subequations}
\begin{align}
\widetilde{E}^{0}_{0}&=E^{0}_{0}=0\\
\widetilde{E}^{0}_{\alpha}&=\Lambda^{\beta}_{\alpha}E^{0}_{\beta}=0\\
\widetilde{E}^{\alpha}_{\beta}&=S^{\alpha}_{\kappa}\Lambda^{\lambda}_{\beta}
E^{\kappa}_{\lambda}=0,
\end{align}
\end{subequations}
where $S^{\alpha}_{\beta}$ is the inverse of
$\Lambda^{\alpha}_{\beta}$.This can be explicitly seen by
observing that the extrinsic curvature transforms as a (0,2)
tensor under these transformations, despite the mixing of time and
space coordinates. The ef\mbox{}fect of a time reparameterization
is trivially seen also to be a covariance . Finally some
terminology is needed; (\ref{quadraticconstraint}) is called
\emph{``Quadratic Constraint''}, (\ref{linearconstraint}) are
called \emph{``Linear Constraints''}, and (\ref{equationofmotion})
are simply the \emph{``Equations of Motion''}.
\section{Essential Constants}
The task of finding the maximal number of essential constants for
each model is complicated by the presence of the quadratic and
linear constraint equations.\\ The first thing to observe is that
their time-derivatives vanish by virtue of the spatial equations
of motion; therefore, they are first integrals of motion for these
equations and they will be satisfied at all times once they are
satisfied at one instant of time. The constraint equations can
thus be considered as algebraic relations restricting the initial
data at some arbitrarily chosen hypersurface. Accordingly, one
initial datum will be absent for each such independent
constraint.\\ The second important thing is that the additive
constant of integration at the right-hand side of these constraint
equations is identically zero. This points to the fact that the
presence of these equations signals the existence of "gauge"
symmetry for the whole system of equations, namely the
time-dependent A.I.D.'s briefly described in the previous section.
Under these transformations, one more constant becomes absorbable.
Thus, if we wish to consider the constraints as full fledged
(first class) dif\mbox{}ferential equations, we have to subtract
two degrees of freedom ( constants in our case) for each such
independent equation.\\ Both points of view are correct and valid:
they are nothing but dif\mbox{}ferent aspects of the same
ingredients of the theory of dif\mbox{}ferential equations. Thus
they should yield the same final result concerning the maximal
number of essential constants. Bellow we present the counting
algorithms of this number for all 30 (4+1) simply transitive,
spatially homogeneous vacuum geometries, according to both points
of view.
 \label{counting}
\subsection{The Initial Value Theorem}
In this section, we apply the initial value theorem to find the
maximal number of essential constants each line element should
contain in order to describe the entire space of solutions for the
given model. In 3+1 dimensions, such a counting has been done some
time ago (see e.g. \cite{Ellis} and the older references therein)
using the Behr decomposition of the structure constants
$C^{\alpha}_{\beta\gamma}$ for 3-dimensional Lie algebras.
However, such a decomposition is not known for Lie algebras of
dimension 4 or higher. Hence, we have to apply an alternative
counting procedure in order to find the essential constants.

A counting which is independent of the dimension can be given using
the initial value theorem. This theorem is stated for the (3+1)-dimensional case
in, for example, Wald's book \cite{Wald}. However, it is fairly easily seen
that this theorem is valid in any dimension; the arguments in the
proof does not depend explicitly on the dimension of the spacetime.

Roughly speaking, the initial value formulation says that a
spacetime satisfying the Einstein equations is uniquely determined
by specifying the metric, $h_{\alpha\beta}$, and the corresponding
extrinsic curvature, $K_{\alpha\beta}$, of an initial spatial
hypersurface (i.e. $\gamma_{\alpha\beta}(t_0)=h_{\alpha\beta}$)
--at the Gauss normal coordinates system, in which the shift
vanishes. The initial data must also satisfy the quadratic
constraint, and the linear constraint on the initial hypersurface
which are purely algebraic in the initial data. Furthermore,
isometric dif\mbox{}feomorphisms on the initial hypersurface, can
always be extended to isometric dif\mbox{}feomorphisms of the
entire spacetime.

The theorem does not mention whether two dif\mbox{}ferent initial
data can lead to the same spacetime. However, any initial data
always generates a one-parameter family of data which will yield
the same maximal development. This one-parameter family is exactly
the time evolution of the pair
($\gamma_{\alpha\beta}(t),K_{\alpha\beta}(t)$). Hence, for a
spacetime foliated into spatial hypersurfaces, any hypersurface
may serve as an initial hypersurface.

The initial value formulation thus provides us with the following
algorithm for counting the essential constants for the spatially
homogeneous model of type $A$: \beq
\#(h_{\alpha\beta},K_{\alpha\beta})-\dim\mathrm{Aut}(A)-\#(\text{independent
constraints})-1. \eeq In our case,
$\#(h_{\alpha\beta},K_{\alpha\beta})=20$, since $h_{\alpha\beta}$
and $K_{\alpha\beta}$ are  symmetric 2-tensors. $\mathrm{Aut}(A)$
is the automorphism group for the Lie algebra $A$; these
automorphisms can be seen to be the ef\mbox{}fect of isomorphic
dif\mbox{}feomorphisms on the initial hypersurface. Thus they
carry the relevant ``gauge'' freedom which must be subtracted
\cite{CMP, ashtekar}. On the initial hypersurface the constraints
(quadratic plus linear constraints) are only algebraic equations,
thus subtract one for each constraint. Finally, we subtract 1 due
to the fact that each initial hypersurface traces out a
one-parameter family of initial data each giving rise to the same
spacetime.

Using the above algorithm we produced Tables \ref{table1} and
\ref{table2} giving the number of essential constants for all 30,
simply transitive, spatially homogeneous vacuum cosmological
models of dimension 4+1. Table \ref{table1} contains the essential
constants for the general form of the algebras, while Table
\ref{table2} contains the essential constants for the exceptional
cases in which for some values of the parameters (of the Lie
algebra) some of the linear constraints vanish identically.

\subsection{Time Dependent A.I.D.s}
In this section, we apply the time dependent A.I.D.s to perform a
second independent counting of the maximal number of essential
constants each line element should contain in order to describe
the entire space of solutions for the given model. This way of
counting is valid in any spatial gauge. The key observation is
that the solutions to the integrability conditions (\ref{Aut})
always contain $4$ arbitrary functions of time. These arbitrary
functions are distributed in $\Lambda^{\alpha}_{\beta}$ and
$P^{\alpha}$ in a way that dif\mbox{}fers for each of the 30
models; e.g. to take an extreme case in the Kasner-like model,
$4A_1$, $\Lambda^{\alpha}_{\beta}$ is completely constant while
all 4 arbitrary functions of time are located in $P^{\alpha}$. In
all cases, $P^{\alpha}$ contains all arbitrary functions through
either their derivatives, or themselves. Thus two distinct ways of
using the gauge freedom suggest themselves, leading to two
versions of the counting algorithm:

The first, is to use the whole freedom in order to set the shift
$\widetilde{N}^{\alpha}$ equal to zero and then see how many first
class linear constraints remain. The corresponding version of the
algorithm is:\\
$D=2\times (\#$ of $\gamma_{\alpha\beta}$)\\
$-2\times \#$ ( linear constraints )\\
$-2\times $ (the Quadratic Constraint)\\
$-\#$ (parameters of Outer Automorphic matrices)\\
$-(\kappa$)\\
where $\kappa\equiv\textrm{dim(Inner)}-\#~\textrm{functionally
independent
Linear Constraints}$.\\
Peano's theorem requires 2 initial data for each
$\gamma_{\alpha\beta}$ since the system is of second order. We
subtract 2 constants for each independent first class constraint.
Finally we subtract the remaining rigid symmetries which are the
parameters of the outer Automorphisms plus the dif\mbox{}ference
between the number of parameters of the inner automorphisms
subgroup and the number of functionally independent linear
constraints.

The second consists of all other options, e.g. we can use the
functions of time contained in $\Lambda^{\alpha}_{\beta}$ to
simplify the scale factor matrix $\gamma_{\alpha\beta}(t)$ and the
remaining functions contained in $P^{\alpha}$ --if any-- to alter
somehow the initial shift vector (e.g. equating some components or
setting some of them equal to zero). Now the algorithm reads:\\
$D=2\times (\#$ of $\gamma_{\alpha\beta})+1\times (\#$ of possibly
remaining\footnote{i.e. after solving algebraically as many as
linear constraints is possible --in terms of the shift's vector
components--, i.e. the shift components which are not expressed in
terms of the scale factor matrix components and their derivatives.} shift vector's components)\\
$-2\times \#$ (of those linear constraints which do not
finally involve shift vector's components)\\
$-2\times $ (the Quadratic Constraint)\\
$-\#$ (parameters of those Outer Automorphic matrices which
preserve the form of the reduced $\gamma_{\alpha\beta}$)\\

For the sake of illustration we give bellow three examples of
counting with both versions of the algorithm presented in this
subsection.

\subsubsection{Type $A_{2}\oplus A_{1}$} The structure constants
are: $C^{2}_{12}=1$. Thus:
\begin{displaymath}
\Lambda^{\alpha}_{\beta}(t)=\left(
\begin{array}{cccc}
  1 & 0 & 0 & 0 \\
  \lambda_{5}(t) & \lambda_{6}(t) & 0 & 0 \\
  \lambda_{9} & 0 & \lambda_{11} & \lambda_{12} \\
  \lambda_{13} & 0 & \lambda_{15} & \lambda_{16} \\
\end{array}
\right)
\end{displaymath}
\begin{displaymath}
P^{\alpha}(t)=\left\{ \frac{\lambda_6'(t)}{2\,\lambda_6(t)},
  \frac{-\left( \lambda_6(t)\,\lambda_5'(t) -
       \lambda_5(t)\,\lambda_6'(t) \right) }
     {2\,\lambda_6(t)},p_{3}(t),
  p_{4}(t)\right\}
\end{displaymath}
Four functions of time appear --as expected; two of them in
$\Lambda^{\alpha}_{\beta}(t)$ and correspond to the inner
automorphism proper invariant subgroup. The number of functionally
independent linear constraints, is 4.
\paragraph{1st version} We
use our entire freedom in order to set the shift vector equal to
zero. So:
\begin{eqnarray*}
\#~\gamma_{\alpha\beta}=10 \\
\#~\textrm{Linear Constraints in terms of}~\dot{\gamma}_{\alpha\beta}=4\\
\#~\textrm{of parameters of Out. Aut. matrices}=6\\
\kappa=2-4=-2\\
D=2\times 10-2\times 4-2-6-(-2)=6
\end{eqnarray*}

\paragraph{2nd version} We use our freedom in order to set:
$N^{3}(t)=N^{4}(t)=0$ and $\gamma_{12}(t)=0$, $\gamma_{22}(t)=1$.
So:
\begin{eqnarray*}
\#~\gamma_{\alpha\beta}=8\\
\#~\textrm{remaining}~N^{\alpha}=0\\
\#~\textrm{Linear Constraints in terms of}~\dot{\gamma}_{\alpha\beta}=2\\
\#~\textrm{of parameters of those Out. Aut. matrices which
preserve the form of}~\gamma_{\alpha\beta}=4\\
D=2\times 8-2\times 2-2-4=6
\end{eqnarray*}
\subsubsection{Type $A_{3,6}\oplus A_{1}$} The structure constants
are: $C^{2}_{13}=-1$ and $C^{1}_{23}=1$. Thus:
\begin{displaymath}
\Lambda^{\alpha}_{\beta}(t)=\left(
\begin{array}{cccc}
  c \cos(f(t)) & c \sin(f(t)) & \lambda_{3}(t) & 0 \\
  -c \sin(f(t)) & c \cos(f(t)) & \lambda_{7}(t) & 0 \\
  0 & 0 & 1 & 0 \\
  0 & 0 & \lambda_{15} & \lambda_{16} \\
\end{array}
\right)
\end{displaymath}
\begin{displaymath}
P^{\alpha}(t)=\left\{ \frac{-\left( \lambda_{3}(t)\,f'(t) \right)  -
     \lambda_{7}'(t)}{2},
  \frac{-\left( \lambda_{7}(t)\,f'(t) \right)  +
     \lambda_{3}'(t)}{2},\frac{-f'(t)}{2},
  p_{4}(t)\right\}
\end{displaymath}
Four functions of time appear --as expected; the three in
$\Lambda^{\alpha}_{\beta}(t)$ and correspond to the inner
automorphism proper invariant subgroup. The number of functionally
independent linear constraints, is 3. \paragraph{1st version} We
use our entire freedom in order to set the shift vector equal to
zero. So:
\begin{eqnarray*}
\#~\gamma_{\alpha\beta}=10\\
\#~\textrm{Linear Constraints in terms of}~\dot{\gamma}_{\alpha\beta}=3\\
\#~\textrm{of parameters of those Out. Aut. matrices
which preserve the form of}~\gamma_{\alpha\beta}=3\\
\kappa=3-3=0\\
D=2\times 10-2\times 3-2-3=9
\end{eqnarray*}

\paragraph{2nd version} We use our freedom in order to set:
$N^{1}(t)=N^{2}(t)=N^{4}(t)=0$ and $\gamma_{12}(t)=0$. So:
\begin{eqnarray*}
\#~\gamma_{\alpha\beta}=9\\
\#~\textrm{remaining}~N^{\alpha}=0\\
\#~\textrm{Linear Constraints in terms of}~\dot{\gamma}_{\alpha\beta}=2\\
\#~\textrm{of parameters of those Out. Aut. matrices
which preserve the form of}~\gamma_{\alpha\beta}=3\\
D=2\times 9-2\times 2-2-3=9
\end{eqnarray*}
\subsubsection{Type $A^{-\frac{1}{3},-\frac{1}{3}}_{4,5}$} The
structure constants are: $C^{1}_{14}=1$, $C^{2}_{24}=-\frac{1}{3}$
and $C^{3}_{34}=-\frac{1}{3}$. Thus:
\begin{displaymath}
\Lambda^{\alpha}_{\beta}(t)=\left(\begin{array}{cccc}
  \lambda_{1}(t) & 0 & 0 & \lambda_{4}(t) \\
  0 & \frac{c1}{\lambda_{1}(t)^{3}} & \frac{c2}{\lambda_{1}(t)^{3}} & \lambda_{8}(t) \\
  0 & \frac{c3}{\lambda_{1}(t)^{3}} & \frac{c4}{\lambda_{1}(t)^{3}} & \lambda_{12}(t) \\
  0 & 0 & 0 & 1 \\
\end{array}\right)
\end{displaymath}
\begin{displaymath}
P^{\alpha}(t)=\left\{ \frac{-\left( \lambda_{4}(t)\,\lambda_{1}'(t) -
       \lambda_{1}(t)\,\lambda_{4}'(t) \right) }{2\,
     \lambda_{1}(t)},\frac{-\left( \lambda_{8}(t)\,
        \lambda_{1}'(t) +
       3\,\lambda_{1}(t)\,\lambda_{8}'(t) \right) }{2\,
     \lambda_{1}(t)},8\leftrightarrow 12,
  \frac{-\lambda_{1}'(t)}{2\,\lambda_{1}(t)}\right\}
\end{displaymath}
Four functions of time appear --as expected; all the four in
$\Lambda^{\alpha}_{\beta}(t)$ and correspond to the inner
automorphism proper invariant subgroup. The number of functionally
independent linear constraints, is 2. \paragraph{1st version} We
use our entire freedom in order to set the shift vector equal to
zero. So:
\begin{eqnarray*}
\#~\gamma_{\alpha\beta}=10\\
\#~\textrm{Linear Constraints in terms of}~\dot{\gamma}_{\alpha\beta}=2\\
\#~\textrm{of parameters of those Out. Aut. matrices
which preserve the form of}~\gamma_{\alpha\beta}=4\\
\kappa=4-2=2\\
D=2\times 10-2\times 2-2-4-2=8
\end{eqnarray*}

\paragraph{2nd version} We use our freedom in order to
set: $\gamma_{11}(t)=1$ and
$\gamma_{14}(t)=\gamma_{24}(t)=\gamma_{34}(t)=0$. So:
\begin{eqnarray*}
\#~\gamma_{\alpha\beta}=6\\
\#~\textrm{remaining}~N^{\alpha}=2\\
\#~\textrm{Linear Constraints in terms of}~\dot{\gamma}_{\alpha\beta}=0\\
\#~\textrm{of parameters of those Out. Aut. matrices
which preserve the form of}~\gamma_{\alpha\beta}=4\\
D=2\times 6+2-2-4=8
\end{eqnarray*}

\section{Exact solutions}
\label{exact}
We will here provide with examples of spatially homogeneous vacuum solutions in 4+1
dimensions\footnote{Some of the solutions are previously known,
even though in many cases the true number of free parameters was not
recognized.}.  There are some general things worth noting. For the decomposable
cases, $A_3\oplus A_1$, we can generate vacuum solutions from
scalar field solutions of the Bianchi models in 3+1 dimensions. More
explicitly, given a
vacuum solution in 4+1 dimensions with metric
\beq
ds_5^2=ds^2_4+e^{-2\phi}\dy^2,
\eeq
the metric $d\tilde{s}_4^2=e^{-\phi}ds_4^2$
will be a solution to the (3+1)-dimensional Einstein equations with a
scalar field. Thus, by going the other way, we can construct vacuum
solutions in 4+1 dimensions from scalar field solutions in one
dimension lower. In many cases (like type VIII$\oplus \mb{R}$ and
IX$\oplus \mb{R}$) these are the only non-trivial solutions one knows
explicitly (see \cite{Kramer}).

The main object of this section is to give some examples of  solutions
of the various types. For only two types we know all the possible
exact vacuum solutions, the remaining cases we only know some special
ones.

\subsection{$4A_1=$I$\oplus \mb{R}$}
There is a 2-parameter family of Kasner solutions which exhaust all
solutions of this type \cite{CD,SH}:
\beq
ds^2=-\dt^2+\sum_{i=1}^4t^{2p_i}\dx^i\dx^i,
\eeq
where $\sum_ip_i=\sum_ip_i^2=1$.

\subsection{$A_2\oplus 2A_1=$III$\oplus \mb{R}$}
There is a 2-parameter family of plane-wave solutions which can be
obtained by restricting the type VI$_h\oplus \mb{R}$ plane-waves
(III=VI$_{-1}$) (see section \ref{VIhR}).

Also, there is a 2-parameter family of solutions, with a higher
symmetry, given by:
\beq
ds^2 &=& -\frac{k^2\omega^2e^{-2(1+a)t}\dt^2}{\sinh^4\omega
t}+\frac{k^2e^{-2(1+a)t}}{\sinh^2\omega
t}\left(\dx^2+e^{-2x}\dy^2\right)
+e^{2at}\dz^2+e^{2t}\dw^2, \\
\omega^2 &=& a^2+a+1. \nonumber
\eeq
The symmetry group of these solutions is
$G_5=SL(2,\mb{R})\times\mb{R}^2$ which acts transitively on the
spatial hypersurfaces.
\subsection{$2A_2$}
There is one solution which can be obtained by a Wick rotation of a
solution in \cite{DeSmet}:
\beq
ds^2=-\dt^2+\frac{t^2}{3}\left[(\dx^2+e^{2x}\dy^2)+(\dz^2+e^{2z}\dw^2)\right].
\eeq
This has indeed the bigger symmetry group $G_6=SL(2,\mb{R})\times SL(2,\mb{R})$ acting on the
spatial hypersurfaces
$\Sigma_t$. It is also algebraically special of type 22 in the sense
of \cite{DeSmet}.
\subsection{$A_{3,1}\oplus A_1=$II$\oplus\mb{R}$}
The general solutions (containing 6 parameters) are not known to our
knowledge, but we have found a 4-parameter family of solutions. \footnote{See also \cite{Halpern} which considers the
$A_{3,1}\oplus A_1$ and $A_{3,3}\oplus A_1$ cases.} It is given by:
\beq
ds^2&=&-\frac{a_4}{\omega}e^{(a_1+a_2+3a_3)t}\cosh\omega
t\dt^2+\frac{\omega}{a_4}\frac{e^{-a_3 t}}{\cosh\omega t}\left(\dx-z\dy\right)^2 \nonumber
\\
 &&+e^{a_3t}\frac{\cosh\omega
t}{\omega}\left(e^{a_1t}\dy^2+e^{a_2t}\dz^2\right)+e^{2a_3
t}\dw^2, \label{sol:IIR}\eeq where
$\omega^2=a_1a_2+2(a_1+a_2)a_3+a_3^2$.

This family of solutions generalizes Taub's type II vacuum
solutions.
\subsection{$A_{3,2}\oplus A_1=$IV$\oplus\mb{R}$}
A 3-parameter family of plane-wave solutions is given by eq.(\ref{sol:A42}) with $s=2\beta_+$.
\subsection{$A_{3,3}\oplus A_1=$V$\oplus\mb{R}$}
A 2-parameter family  of plane-wave solutions is given by eq.(\ref{sol:A45}) with
$s=2\beta_+$.

There is also 3-parameter family of solutions given by:
\beq
ds^2&=&-\frac{k^2\omega^2e^{-a_1t}\dt^2}{4\sinh^3\omega t}+e^{2a_1
t}\dw^2 \nonumber \\ && +
\frac{k^2e^{-a_1 t}}{\sinh\omega t}\left(e^{a_2 t}e^{-2z}\dx^2+e^{-a_2
t}e^{-2z}\dy^2+\dz^2\right), \\
3\omega^2 &=& 3a_1^2+a_2^2. \nonumber
\eeq
\subsection{$A_{3,4}\oplus A_1=$VI$_0\oplus\mb{R}$}
There is a 1-parameter family of solutions given by
eq. (\ref{sol:A450}) with $p=-1$, and $q=0$.
\subsection{$A_{3,5}^p\oplus A_1=$VI$_h\oplus\mb{R}$}
\label{VIhR}
A 3-parameter family  of plane-wave solution is given by
eq.(\ref{sol:A45}) with $s=2\beta_+$. There are also a 2-parameter
family of solutions
given by eq. (\ref{sol:A45DM}) with $q=0$.
\subsection{$A_{3,6}\oplus A_1=$VII$_0\oplus\mb{R}$}
Apart from the solutions with higher symmetry deducible from scalar
field Bianchi type VII$_0$, the authors do not know of any other non-trivial
solutions.
\subsection{$A_{3,7}^p\oplus A_1=$VII$_h\oplus\mb{R}$}
A 3-parameter family  of plane-wave solution is given in eq.(\ref{sol:A46}) with
$s=2\beta_+$.
\subsection{$A_{3,8}\oplus A_1=$VIII$\oplus\mb{R}$}
Apart from the solutions with higher symmetry deducible from scalar
field Bianchi type VIII, the authors do not know of any other non-trivial
solutions (see also \cite{LP86}).
\subsection{$A_{3,9}\oplus A_1=$IX$\oplus\mb{R}$}
Apart from the solutions with higher symmetry deducible from scalar
field Bianchi type IX, the authors do not know of any other non-trivial
solutions (see also \cite{LP86}).

\subsection{$A_{4,1}$}
No vacuum solutions of this type is known to the authors.\footnote{There are some
known self-similar solutions with a perfect fluid \cite{sig4+1}. Note
there is a typo in eq. (83); all exponents should be divided by $\gamma$.}

\subsection{$A_{4,2}^p$}
There is a 3-parameter family of plane-wave solutions for $p> -2$ \cite{sigpp}:
\beq
ds^2 =&& e^{2t}(-\dt^2+\dw^2) + e^{2s(w+t)}\nonumber \\
&\times&\bigg[e^{-4\beta_+(w+t)}\left(\dx+\frac{Q_1}{P_1}e^{3\beta_+(w+t)}\dy+\left[A+B(w+t)\right]e^{3\beta_+(w+t)}\dz\right)^2
\nonumber \\
&+&
e^{2\beta_+(w+t)}\left(\dy+{Q_3}(w+t)\dz\right)^2+e^{2\beta_+(w+t)}\dz^2\bigg],
\label{sol:A42}\eeq
where
\beq
s(1-s)&=& 2\beta_+^2+\frac{1}{6}(Q_1^2+Q_2^2+Q_3^2)\nonumber \\
P_1 &=& 3\beta_+,
\label{eq:P1}\eeq
and
\beq
A  =  \frac{3\beta_+Q_2-Q_1Q_3}{3\beta_+},\quad
B  =  \frac{Q_1Q_3}{3\beta_+}.
\label{eqAB}\eeq
The group parameter is given by:
\beq
p=\frac{s-2\beta_+}{s+\beta_+}.
\eeq

For $p=-2$, there is a 1-parameter family of vacuum solutions due to
Demaret and Hanquin\footnote{However, they did not realize that the
solution was part of a one-parameter family of solutions.} \cite{DH}:
\beq
ds^2 &=& k^2e^{3t^2}t^{-\frac{1}{24}}\left(t^{-\frac 12}+t^{\frac 12}\right)\left(-\dt^2+\dw^2\right)+t^{\frac 23}e^{4w}\dx^2\nonumber
\\
&& +t^{\frac 53}(t^{-\frac 12}+t^{\frac
12})e^{-2w}\dy^2+\frac{t^{-\frac 13}}{t^{-\frac 12}+t^{\frac 12}}e^{-2w}(\dz-w\dy)^2.
\eeq
\subsection{$A_{4,2}^1$}
There is a 2-parameter family of plane-wave solutions of one sets $Q_1=0$, and then $\beta_+=0$ in eq. (\ref{sol:A42}).
\subsection{$A_{4,3}$}
Plane-wave solutions for the Lie algebra type $A_{4,3}$ can be
obtained by taking the $p\longrightarrow\infty$ limit of
$A_{4,2}^p$. In this limit we get $\beta_+=-s$ and thus the metric can
be written \cite{sigpp}:
\beq
ds^2 =&& e^{2t}(-\dt^2+\dw^2) \nonumber \\
&+&e^{6s(w+t)}\left(\dx+\frac{Q_1}{P_1}e^{-3s(w+t)}\dy+\left[A+B(w+t)\right]e^{-3s(w+t)}\dz\right)^2
\nonumber \\
&+&\left(\dy+{Q_3}(w+t)\dz\right)^2
+\dz^2,
\eeq
where
\beq
s=\frac 16\left(1\pm\sqrt{1-2(Q_1^2+Q_2^2+Q_3^2)}\right),
\eeq
and $A,B$ are given in eq. (\ref{eqAB}) with $\beta_+=-s$.

\subsection{$A_{4,4}$}
There is a 3-parameter set of plane-wave
solutions given by \cite{sigpp}:
\beq
ds^2 =&& e^{2t}(-\dt^2+\dw^2) + e^{2s(w+t)}\nonumber \\
&\times&\bigg[\left(\dx+{Q_1}(w+t)\dy+(w+t)\left[Q_2+\frac{Q_1Q_3}{2}(w+t)\right]\dz\right)^2
\nonumber \\
&+&\left(\dy+{Q_3}(w+t)\dz\right)^2 +\dz^2\bigg],
\eeq
where
\[ s(1-s)=\frac{1}{6}(Q_1^2+Q_2^2+Q_3^2). \]

\subsection{$A^{pq}_{4,5}$}

Given $p+q+1> 0$, a 3-parameter set of plane-wave solutions can
be given by  \cite{sigpp}:
\beq
ds^2 =&& e^{2t}(-\dt^2+\dw^2) + e^{2s(w+t)}\nonumber \\
&\times&\bigg[e^{-4\beta_+(w+t)}\left(\dx+\frac{Q_1}{P_1}e^{P_1(w+t)}\dy+\frac{Q_1Q_3+P_3Q_2}{P_3P_2}e^{P_2(w+t)}\dz\right)^2
\nonumber \\
&+&
e^{2(\beta_++\sqrt{3}\beta_-)(w+t)}\left(\dy+\frac{Q_3}{P_3}e^{P_3(w+t)}\dz\right)^2+e^{2(\beta_+-\sqrt{3}\beta_-)(w+t)}\dz^2\bigg],
\label{sol:A45}\eeq
where
\beq
&s(1-s)= 2(\beta_+^2+
\beta_-^2)+\frac{1}{6}(Q_1^2+Q_2^2+Q_3^2)\nonumber \\
&P_1 = 3\beta_++\sqrt{3}\beta_-,\quad
P_2 = 3\beta_+-\sqrt{3}\beta_-, \quad
P_3 = -2\sqrt{3}\beta_-.
\label{eq:Ps}\eeq
The group parameters are related to these parameters as follows:
\beq
p = \frac{s + (\beta_++\sqrt{3}\beta_-)}{s+(\beta_+-\sqrt{3}\beta_-)},\quad
q = \frac{s-2\beta_+}{s + (\beta_+-\sqrt{3}\beta_-)}.
\eeq

There are also some other solutions due to Demaret and Hanquin
\cite{DH}\footnote{They only give it as a 1-parameter family.}. Given
$p+q+1\neq 0$, then there is a 2-parameter family of vacuum solutions:
\beq
ds^2=k^2(\sinh t)^{2\sum P_i^2}\left(\tanh\frac t2\right)^{2\sum \alpha_iP_i}\left(-\dt^2+\dw^2\right) \nonumber \\
+\sum(\sinh t)^{2P_i}\left(\tanh\frac t2\right)^{2\alpha_i}e^{2P_iw}\left(\dx^i\right)^2,
\label{sol:A45DM}\eeq
where $\sum P_i=1$, $\sum\alpha_i=0$, and $\sum\alpha_i^2=1+\sum
P_i^2$.

Given $p+q+1=0$, there is a 1-parameter family of  solutions due to
Demaret and Hanquin \cite{DH}:
\beq
ds^2&=& k^2e^{(1+p^2+q^2)\frac{t^2}{2}}t^{-\frac 23}(-\dt^2+\dw^2)+t^{\frac 23}\left(e^{2w}\dx^2+e^{2pw}\dy^2+e^{2qw}\dz^2\right).
\label{sol:A450}\eeq

Also, for the exceptional case $A^{pq*}_{4,5}$ ($q=-(1+p)/2$), we have found a
self-similar solution which generalizes the Collinson-French type
VI$^*_{-1/9}$ vacuum:
\beq
ds^2&=&-{\bf dt}^2+t^2{\bf
dx}^2+\left[t^{\frac{(1-p)^2}{b}}\exp\left({-\sqrt{6}(1+p)rx}\right){\bf
dy}+\frac{1}{2r\sqrt{b}}t{\bf dx}\right]^2\nonumber
\\
&&+t^{\frac{6(1+p)}{b}}\exp\left({4\sqrt{6}rx}\right){\bf
dz}^2+t^{\frac{6p(1+p)}{b}}\exp\left({4p\sqrt{6}rx}\right){\bf dw}^2
\eeq
where
\beq
r = \frac{\sqrt{1+p+p^2}}{5p^2+2p+5}, \quad b = 5p^2+2p+5.
\eeq

\subsection{$A_{4,5}^{p,p}$}
This is a special case of the above. For the plane-wave
solutions, eq. (\ref{sol:A45}),
one has to set $Q_1=0$, and then $P_1=0$.

\subsection{$A_{4,5}^{p,1}$}
Similarly as in the above case, but now set $Q_2=0$ and then $P_2=0$
in eq. (\ref{sol:A45}).

In addition to this, we have found a 3-parameter family of solutions
for the particular value $p=-1$. It is given by:
\beq
ds^2 &=&-\frac{\omega^2k^2e^{-(4a_1+2a_2)t}\dt^2}{\sinh^8\omega
t}+\frac{e^{-a_1t}e^{-2w}}{\sinh^2\omega
t}(e^{-a_2t}\dx^2+\dz^2)\nonumber \\&& +
e^{(2a_1+a_2)t}e^{2w}\sinh^2\omega
t\dy^2+\frac{k^2e^{-(4a_1+2a_2)t}\dw^2}{\sinh^6\omega t}, \\
 8\omega^2&=& 3a_1^2+3a_1a_2+a_2^2. \nonumber
\eeq
\subsection{$A_{4,5}^{1,1}$}
The whole set of solutions is in this case known\footnote{All
solutions with a $\gamma$-law non-tilted perfect fluid is also known,
see \cite{sig4+1}.}. The set is 2-dimensional
and the general solution is given by eq. (\ref{sol:A45DM}) with the
restriction $P_1=P_2=P_3=1/3$. Explicitly,
\beq
ds^2&=&k^2\sinh^{\frac 23}t\left(-\dt^2+\dw^2\right)\nonumber \\
&&+\sinh^{\frac 23}te^{\frac 23w}\left[\left(\tanh\frac
t2\right)^{2a_1}\dx^2+\left(\tanh\frac
t2\right)^{2a_2}\dy^2+\left(\tanh\frac
t2\right)^{-2(a_1+a_2)}\dz^2\right],\nonumber  \\
2 &=& 3a_1^2+3a_2^2+3a_1a_2.
\eeq

\subsection{$A_{4,6}^{pq}$}
Again we have plane-wave solutions \cite{sigpp}:
Let $s$ be
given by
\beq
s(1-s)=2\beta_+^2+\frac 23\omega^2\sinh^22\beta+\frac 16(Q_1^2+Q_2^2).
\eeq
Define also the two one-forms:
\beq
{\mbold\omega}^2&=& \cos[\omega(w+t)] {\bf dy}-\sin[\omega(w+t)]
{\bf dz} \nonumber \\
{\mbold\omega}^3&=& \sin[\omega(w+t)] {\bf dy}+\cos[\omega(w+t)] {\bf dz}.
\eeq
The plane-wave solutions of type $A_{4,6}^{pq}$ can now be written:
\beq
ds^2 =&& e^{2t}(-\dt^2+\dw^2)+e^{2s(w+t)} \nonumber \\
&\times&
\bigg[e^{-4\beta_+(w+t)}\left\{\dx+e^{3\beta_+(w+t)}\left(q_1e^{-\beta}{\mbold\omega}^3-q_2e^{\beta}{\mbold\omega}^2\right)\right\}^2
\nonumber \\
&+&e^{2\beta_+(w+t)}\left\{
e^{-2\beta}\left({\mbold\omega}^2\right)^2+e^{2\beta}\left({\mbold\omega}^3\right)^2\right\}\bigg]
\label{sol:A46}\eeq
where
\beq
q_1 = \frac{Q_1\omega+3\beta_+Q_2e^{2\beta}}{\omega^2+9\beta_+^2}
,\quad
q_2 = \frac{Q_2\omega-3\beta_+Q_1e^{-2\beta}}{\omega^2+9\beta_+^2}.
\eeq
The group parameters are related to these constants via
\beq
p = \frac{\beta_+(s-2\beta_+)}{\omega(s+\beta_+)},\quad
q = \frac{\beta_+}{\omega}.
\eeq
\subsection{$A_{4,7}$}
No such solutions are known to the authors.
\subsection{$A_{4,8}$}
There is a solution which is the $p\rightarrow -1$ limit of the
metric (\ref{sol:A49}).
\subsection{$A_{4,9}^p$}
There is a simple power-law solution for each $-1<p\leq 1$:
\beq
ds^2&=&-\dt^2+t^2\dw^2+k^2t^{\frac{2(2p^2+5p+2)}{3(p^2+p+1)}}e^{-2(p+1)\sigma w}(\dx-z\dy)^2 \nonumber \\
&&+t^{\frac{2(p+2)^2}{3(p^2+p+1)}}e^{-2\sigma w}\dy^2  +t^{\frac{2(2p+1)^2}{3(p^2+p+1)}}e^{-2p\sigma w}\dz^2,
\label{sol:A49}\eeq
where
\beq
\sigma^2=\frac{7p^2+13p+7}{6(p^2+p+1)^2},\quad k^2=\frac{2(7p^2+13p+7)}{9(p^2+p+1)}.
\eeq
Due to the power-law dependence, this solution is self-similar.

There is one special case worth noting, namely $p=-1/2$. \footnote{This
  corresponds to a Lie algebra acting simply transitive on the model geometry $\mb{F}^4$ \cite{sigEssay}.} In this case
the metric can be written:
\beq
ds^2&=&-\dt^2+\frac 32t^2\left(\dw^2+e^{-2w}\dy^2\right)+e^{-w}(\dx-z\dy)^2+e^w\dz^2.
\eeq
Note that the spatial hypersurfaces are fiberbundles over
$\mb{H}^2$. In fact, the symmetry group is larger for this metric than
one would expect; it is the
semi-direct product $G_5=\mb{R}^2\ltimes SL(2,\mb{R})$ with a $U(1)$
stabilizer.

\subsection{$A_{4,9}^1$}
There is a solution obtained from eq. (\ref{sol:A49}) by setting
$p=1$, which is fairly interesting \cite{sigEssay}. By a rescaling  of the coordinates
the solution can be written:
\beq
ds^2&=&-\dt^2 \nonumber \\
&+&\frac{t^2}{2}\left[\dw^2+e^{-2w}\left(\dx+\frac 12(y\dz-z\dy)\right)^2+e^{-w}(\dy^2+\dz^2)\right].
\label{sol:HC2}\eeq
In this case the spatial surfaces are isometric to the complex
hyperbolic space, $\mb{H}_{\mb C}^2$, and hence, it has an
8-dimensional isometry group, $G_8=PU(2,1)$, acting multiply
transitive on the spatial surfaces (it has a $U(2)$ stabilizer).
\subsection{$A_{4,9}^0$}
There is a solution obtained from eq. (\ref{sol:A49}) by setting
$p=0$.
\subsection{$A_{4,10}$}
This algebra acts simply transitive on ${\sf Nil}^3\times{\mb{R}}$ \cite{sigEssay}, so all
solutions of the type II$\oplus \mb{R}$ admitting an extra symmetry
acting on the spatial surfaces, are also invariant under this
algebra. Hence, the solutions (\ref{sol:IIR}) with $a_1=a_2$ are
invariant under this group. Solutions with $A_{4,10}$ as a maximal
symmetry are not known to the authors.
\subsection{$A_{4,11}^p$}
The solution (\ref{sol:HC2}) is invariant under this group due to the
fact that this algebra acts simply transitive on $\mb{H}_{\mb
C}^2$ \cite{sigEssay}. Other solutions are not known to the authors.
\subsection{$A_{4,12}$}
This algebra acts simply transitive on $\mb{H}^3\times \mb{R}$
\cite{sigEssay} so all solutions having this higher symmetry group
are invariant under $A_{4,12}$. An interesting example -- although
far from general -- is the 1-parameter family of solutions
obtained by Wick-rotating the 5D Schwarzschild solution: \beq
ds^2=-\frac{t^2\dt^2}{t^2+2M}+\frac{1}{t^2}\left(t^2+2M\right)\dx^2+t^2\left[\dy^2+e^{-2y}(\dz^2+\dw^2)\right].
\eeq As is clearly seen, this solution has a far larger symmetry
group than $A_{4,12}$, namely $G_7=SL(2,\mb{C})\times \mb{R}$.
However, solutions with a maximal symmetry group $A_{4,12}$ are
not known to the authors.
\begin{center}
\begin{table}
\begin{tabular}{|c|c|}
\multicolumn{2}{c}{\textbf{\textit{TABLE}}}\\
  \hline
  \hline
  Lie Algebra & Non Vanishing Structure Constants\\
  \hline
  \hline
 $4A_{1}$ &  \\
  \hline
  $A_{2}\oplus A_{1}$ & $C^{2}_{12}=1$\\
  \hline
  $2A_{2}$ & $C^{2}_{12}=1$ $C^{4}_{34}=1$\\
\hline
  $A_{3,1}\oplus A_{1}$ & $C^{1}_{23}=1$\\
  \hline
$A_{3,2}\oplus A_{1}$ & $C^{1}_{13}=1$ $C^{1}_{23}=1$
$C^{2}_{23}=1$ \\
\hline $A_{3,3}\oplus A_{1}$ & $C^{1}_{13}=1$ $C^{2}_{23}=1$\\
\hline $A_{3,4}\oplus A_{1}$ & $C^{1}_{13}=1$ $C^{2}_{23}=-1$\\
\hline
  $A^{\alpha}_{3,5}\oplus A_{1}$ $0<|\alpha|<1$ & $C^{1}_{13}=1$ $C^{2}_{23}=\alpha$\\
  \hline
  $A_{3,6}\oplus A_{1}$ & $C^{2}_{13}=-1$ $C^{1}_{23}=1$\\
  \hline
  $A^{\alpha}_{3,7}\oplus A_{1}$ $0<\alpha$ & $C^{1}_{13}=\alpha$ $C^{2}_{13}=-1$ $C^{1}_{23}=1$ $C^{2}_{23}=\alpha$\\
  \hline
  $A_{3,8}\oplus A_{1}$ & $C^{1}_{23}=1$ $C^{2}_{13}=-1$ $C^{3}_{12}=-1$\\
  \hline
  $A_{3,9}\oplus A_{1}$ & $C^{3}_{12}=1$ $C^{1}_{23}=1$ $C^{2}_{31}=1$\\
  \hline
  $A_{4,1}$ & $C^{1}_{24}=1$ $C^{2}_{34}=1$\\
  \hline
  $A^{\alpha}_{4,2}$ $\alpha\neq(0,1)$ & $C^{1}_{14}=\alpha$ $C^{2}_{24}=1$ $C^{2}_{34}=1$ $C^{3}_{34}=1$\\
  \hline
  $A^{1}_{4,2}$ & $C^{1}_{14}=1$ $C^{2}_{24}=1$ $C^{2}_{34}=1$ $C^{3}_{34}=1$\\
\hline $A_{4,3}$ & $C^{1}_{14}=1$ $C^{2}_{34}=1$\\
\hline $A_{4,4}$ & $C^{1}_{14}=1$ $C^{1}_{24}=1$ $C^{2}_{24}=1$
$C^{2}_{34}=1$ $C^{3}_{34}=1$\\
\hline $A^{\alpha,\beta}_{4,5}$ $-1\leq\alpha<\beta<1$,
$\alpha\beta\neq 0$ & $C^{1}_{14}=1$ $C^{2}_{24}=\alpha$
$C^{3}_{34}=\beta$\\
\hline $A^{\alpha,\alpha}_{4,5}$ $-1\leq\alpha<1$, $\alpha\neq 0$
&
$C^{1}_{14}=1$ $C^{2}_{24}=\alpha$ $C^{3}_{34}=\alpha$\\
\hline $A^{\alpha,1}_{4,5}$ $-1\leq\alpha<1$, $\alpha\neq 0$ &
$C^{1}_{14}=1$ $C^{2}_{24}=\alpha$ $C^{3}_{34}=1$\\
\hline $A^{1,1}_{4,5}$ & $C^{1}_{14}=1$ $C^{2}_{24}=1$
$C^{3}_{34}=1$\\
\hline $A^{\alpha,\beta}_{4,6}$ $\alpha\neq 0$, $\beta\geq 0$  &
$C^{1}_{14}=\alpha$ $C^{2}_{24}=\beta$ $C^{3}_{24}=-1$
$C^{2}_{34}=1$ $C^{3}_{34}=\beta$\\
\hline $A_{4,7}$ & $C^{1}_{14}=2$ $C^{2}_{24}=1$ $C^{2}_{34}=1$
$C^{3}_{34}=1$ $C^{1}_{23}=1$\\
\hline
$A_{4,8}$ & $C^{1}_{23}=1$ $C^{2}_{24}=1$ $C^{3}_{34}=-1$ \\
\hline $A^{\beta}_{4,9}$ $0<|\beta|<1$ & $C^{1}_{23}=1$
$C^{1}_{14}=1+\beta$ $C^{2}_{24}=1$ $C^{3}_{34}=\beta$\\
\hline $A^{1}_{4,9}$ & $C^{1}_{23}=1$ $C^{1}_{14}=2$
$C^{2}_{24}=1$ $C^{3}_{34}=1$ \\
\hline $A^{0}_{4,9}$ & $C^{1}_{23}=1$ $C^{1}_{14}=1$
$C^{2}_{24}=1$ \\
\hline
$A_{4,10}$ & $C^{1}_{23}=1$ $C^{3}_{24}=-1$ $C^{2}_{34}=1$\\
\hline $A^{\alpha}_{4,11}$ $\alpha>0$ & $C^{1}_{23}=1$
$C^{1}_{14}=2\alpha$ $C^{2}_{24}=\alpha$ $C^{3}_{24}=-1$
$C^{2}_{34}=1$ $C^{3}_{34}=\alpha$\\
\hline $A_{4,12}$ & $C^{1}_{13}=1$ $C^{2}_{23}=1$ $C^{2}_{14}=-1$
$C^{1}_{24}=1$\\
\hline
\end{tabular}
\caption{The structure constants of all 4-dim, real, Lie Algebras}
\end{table}
\end{center}
\section{Conclusion}
\label{conclusion} We have shown that the usage of the
automorphism group is a very ef\mbox{}ficient way of identifying
the true gravitational degrees of freedom for a simply transitive
spatially homogeneous vacuum geometry. Many investigations have
suf\mbox{}fered from the failure of identifying these. In
particular, if we wish to find the \emph{general} solution under a
given set of assumptions, then it is essential to \emph{ab initio}
identify the number of true degrees of freedom. At this point is
we deem as appropriate to state that the Time-Dependent A.I.D.s
were not only used to derive the second counting algorithm, but
also to find some of the solutions exhibited in section 4.

In this paper we specifically used this method to find the
dimension of the set of all Ricci-flat spatially homogeneous
models of dimension 4+1. Our main results are given in tables
\ref{table1} and \ref{table2}.

Inspecting  tables \ref{table1} and \ref{table2} it is seen that the
most general types have 11 essential constants. Hence, in order to
specify a certain solution under the above assumptions, we need to
specify up to 11 parameters. The maximal number of parameters happens
for the following two types
\[
A_{3,8}\oplus A_1=\textrm{VIII}\oplus\mb{R}\qquad
A_{3,9}\oplus A_1=\textrm{IX}\oplus\mb{R}.
\]
Interestingly, these two  algebras are the trivial extensions of
the Bianchi type Lie algebras VIII and IX and not some
indecomposable ones --as one might have expected. It is also
noteworthy that, the set of the allowed numbers of the Essential
Constants does not contain the numbers 1,3,4 and 5. This does not
occur in 3+1 dimensions where the various models saturate all the
range of values between 1 and 4. There, the models with the
minimum number of essential constants are the Kasner (Type I) and
Joseph (Type V). The corresponding 4+1 counterpart of Type I, i.e.
$4A_{1}$ algebra is seen --by means of the algorithm-- to contain
2 essential constants. Thus why the number 1 is excluded. In fact
this ``hole'' increases with the dimension, since the
corresponding abelian types, will have d-2 essential constants, in
d+1 dimensions. On the other hand, the 4+1 counterparts of the
next ``minimal'' 3+1 models (Type V and its ``neighbour'' Type II
with 1 and 2 essential constants respectively) i.e. the algebras
$A_{3,3}\oplus A_{1}$ and $A_{3,1}\oplus A_{1}$ have both 6
essential constants. The reason for this is that the number of the
"would be constants" depend not only on the more components of the
scale factor matrix $\gamma_{\alpha\beta}(t)$ but also on the
number of the linear constraints (the last being depended on the
algebra). Thus from 2 the number of essential constants is lifted
up to 6. Thus why the numbers between them i.e. 3,4,5 are also
excluded. This sort of ``irregularity'' does not obtain for the
rest of the cases, and thus all the numbers from 6 to 11 appear.

We have also given some exact solutions, some of which are
believed to be new. Only in two cases ($4A_1$ and $A^{1,1}_{4,5}$)
the posited line element is the most general one. For the
remaining types only special solutions are known. However, some of
them -- like the self-similar ones -- may serve as asymptotes for
more general solutions (this does, however, require a stability
analysis within the class under consideration which to date is
only done for the plane-wave solutions \cite{HKL}).
\begin{table}
\begin{tabular}{|c|c|c|c|c|}
\multicolumn{5}{c}{\textbf{\textit{TABLE 1}}}\\
  \hline
  \hline
  Lie Algebra &$\#(h_{\alpha\beta},K_{\alpha\beta})$ & \# of  independent  & $\dim\mathrm{Aut}(A)$ & Essential\\
              & & Linear
  Constraints &   & Constants \\
  \hline
  \hline
  $4A_{1}$ & 20 & 0 & 16 & 2 \\
  \hline
  $A_{2}\oplus 2A_{1}$ & 20 & 4 & 8 & 6 \\
  \hline
  $2A_{2}$ & 20 & 4 & 4 & 10 \\
  \hline
  $A_{3,1}\oplus A_{1}$ & 20 & 2 & 10 & 6 \\
  \hline
  $A_{3,2}\oplus A_{1}$ & 20 & 4 & 6 & 8 \\
  \hline
  $A_{3,3}\oplus A_{1}$ & 20 & 4 & 8 & 6 \\
  \hline
  $A_{3,4}\oplus A_{1}$ & 20 & 3 & 6 & 9 \\
  \hline
  $A^{\alpha}_{3,5}\oplus A_{1}$, $0<|\alpha|<1$ & 20 & 4 & 6 & 8 \\
  \hline
  $A_{3,6}\oplus A_{1}$ & 20 & 3 & 6 & 9 \\
  \hline
  $A^{\alpha}_{3,7}\oplus A_{1}$, $0<\alpha$ & 20 & 4 & 6 & 8 \\
  \hline
  $A_{3,8}\oplus A_{1} ~~~\textrm{and} ~~~A_{3,9}\oplus A_{1}$ & 20 & 3 & 4 & 11\\
  \hline
  $A_{4,1}$ & 20 & 3 & 7 & 8 \\
  \hline
  $A^{\alpha}_{4,2}$, $\alpha\neq \{0,1\}$ & 20 & 4 & 6 & 8 \\
  \hline
  $A^{1}_{4,2}$ & 20 & 4 & 8 & 6\\
  \hline
  $A_{4,3}$ & 20 & 4 & 6 & 8 \\
  \hline
  $A_{4,4}$ & 20 & 4 & 6 & 8 \\
  \hline
  $A^{\alpha,\beta}_{4,5}$, $\alpha,\beta \in [-1,1)-\{0\}$, $\alpha\neq \beta$ & 20 & 4 & 6 & 8\\
  \hline
  $A^{\alpha,\alpha}_{4,5}$, $\alpha \in [-1,1)-\{0\}$ & 20 & 4 & 8 & 6\\
  \hline
  $A^{\alpha,1}_{4,5}$, $\alpha \in [-1,1)-\{0\}$ & 20 & 4 & 8 & 6\\
  \hline
  $A^{1,1}_{4,5}$ & 20 & 4 & 12 & 2\\
  \hline
  $A^{\alpha,\beta}_{4,6}$, $\alpha\neq 0, \beta\geq0$ & 20 & 4 & 6 & 8\\
  \hline
  $A_{4,7}$ & 20 & 4 & 5 & 9\\
  \hline
  $A_{4,8}$ & 20 & 3 & 5 & 10\\
  \hline
  $A^{\beta}_{4,9}$, $0<|\beta|<1$ & 20 & 4 & 5 & 9\\
  \hline
  $A^{1}_{4,9}$ & 20 & 4 & 7 & 7\\
  \hline
  $A^{0}_{4,9}$ & 20 & 4 & 5 & 9\\
  \hline
  $A_{4,10}$ & 20 & 3 & 5 & 10 \\
  \hline
  $A^{\alpha}_{4,11}$, $0<\alpha$ & 20 & 4 & 5 & 9\\
  \hline
  $A_{4,12}$ & 20 & 4 & 4 & 10\\
  \hline
\end{tabular}
\caption{Essential Constants of 4+1 Spatially Homogeneous Models}
\label{table1}
\end{table}
\begin{table}
\begin{tabular}{|c|c|c|c|c|}
\multicolumn{5}{c}{\textbf{\textit{TABLE 2}}}\\
  \hline
  \hline
  Lie Algebra &$\#(h_{\alpha\beta},K_{\alpha\beta})$ & \# of  independent  & $\dim\mathrm{Aut}(A)$ & Essential\\
              & & Linear
  Constraints &   & Constants \\
  \hline
  \hline
  $A^{\alpha}_{3,5}\oplus A_{1}$, $0<|\alpha|<1$& 20 & 3 & 6 & 9 \\
  for $\alpha=-1/2$, & & & &\\
  \hline
  $A^{\alpha}_{4,2}$, for $\alpha=-1,-3$ & 20 & 3 & 6 & 9 \\
  \hline
  $A^{\alpha,\beta}_{4,5}$, $\alpha,\beta \in [-1,1)-\{0\}$ & 20 & 3 & 6 & 9\\
  for $1+2\alpha+\beta=0$ & & & &\\
  or $1+2\beta+\alpha=0$ & & & &\\
  \hline
  $A^{\alpha,\alpha}_{4,5}$, $\alpha \in [-1,1)-\{0\}$& 20 & 3 & 8 & 7\\
  for $\alpha=-1$ & & & &\\
  \hline
  $A^{\alpha,\alpha}_{4,5}$, $\alpha \in [-1,1)-\{0\}$& 20 & 2 & 8 & 8\\
  for $\alpha=-1/3$ & & & &\\
  \hline
  $A^{\alpha,1}_{4,5}$, $\alpha \in [-1,1)-\{0\}$& 20 & 3 & 8 & 7\\
  for $\alpha=-1$ & & & &\\
  \hline
  $A^{\alpha,\beta}_{4,6}$, $\alpha\neq 0, \beta\geq0$& 20 & 3 & 6 & 9\\
  for $\alpha=-\beta$ & & & &\\
  \hline
\end{tabular}
\caption{Essential Constants of 4+1 Spatially Homogeneous
Exceptional Models}
\label{table2}
\end{table}
\vspace*{0.5cm}
\section*{Acknowledgements}
G.O. Papadopoulos is currently a scholar of the Greek State
Scholarships Foundation (I.K.Y.) and acknowledges the relevant
financial support. T Christodoulakis and G.O. Papadopoulos,
acknowledge support by the University of Athens, Special Account
for the Research Grant-No. 70/4/5000. S. Hervik acknowledges funding
from the Research Council of Norway and an Isaac Newton Studentship.

\end{document}